\documentclass[12pt]{article}
\usepackage{amssymb,amsfonts}
\usepackage{graphics,psboxit,amsmath}
\usepackage{subfigure}
\usepackage{graphicx}
\usepackage{verbatim}
\usepackage{color}
\usepackage{hyperref}
\hypersetup{colorlinks}

\definecolor{darkred}{rgb}{1,0,0}
\definecolor{darkgreen}{rgb}{0,0.5,0}
\definecolor{darkblue}{rgb}{0,0,1}
\definecolor{orange}{rgb}{1,0.5,0}
\definecolor{green}{rgb}{0,1,0}
\definecolor{purple}{rgb}{.5,0,1}

\hypersetup{ colorlinks,
linkcolor=darkred,
filecolor=darkgreen,
urlcolor=darkblue,
citecolor=darkblue,
linktocpage=true }

\definecolor{markcolor}{rgb}{.25,0,1}

\definecolor{markcolor2}{rgb}{1,0,0}

\definecolor{markcolor3}{rgb}{0,1,0}

\def\hybrid{\topmargin -10pt    \oddsidemargin 0.1in %%%%%%%%%%%%%% Archive-30pt
        \headheight 0pt \headsep 0pt
        \textwidth 16.0cm      % A4 paper
        \textheight 22.2cm       % A4 paper
        \marginparwidth .875in
        \parskip 5pt plus 1pt   \jot = 1.5ex}

\hybrid

\catcode`\@=11

\def\marginnote#1{}
\newcount\hour
\newcount\minute
\newtoks\amorpm
\hour=\time\divide\hour by60 \minute=\time{\multiply\hour by60
\global\advance\minute by-\hour}
\edef\standardtime{{\ifnum\hour<12 \global\amorpm={am}%
        \else\global\amorpm={pm}\advance\hour by-12 \fi
        \ifnum\hour=0 \hour=12 \fi
        \number\hour:\ifnum\minute<10 0\fi\number\minute\the\amorpm}}
\edef\militarytime{\number\hour:\ifnum\minute<10 0\fi\number\minute}
\def\draftlabel#1{{\@bsphack\if@filesw {\let\thepage\relax
   \xdef\@gtempa{\write\@auxout{\string
      \newlabel{#1}{{\@currentlabel}{\thepage}}}}}\@gtempa
   \if@nobreak \ifvmode\nobreak\fi\fi\fi\@esphack}
        \gdef\@eqnlabel{#1}}
\def\@eqnlabel{}
\def\@vacuum{}
\def\draftmarginnote#1{\marginpar{\raggedright\scriptsize\tt#1}}

\def\draft{\oddsidemargin -.5truein
        \def\@oddfoot{\sl preliminary draft \hfil
        \rm\thepage\hfil\sl\today\quad\militarytime}
        \let\@evenfoot\@oddfoot \overfullrule 3pt
        \let\label=\draftlabel
        \let\marginnote=\draftmarginnote
   \def\@eqnnum{(\theequation)\rlap{\kern\marginparsep\tt\@eqnlabel}%
\global\let\@eqnlabel\@vacuum}  }

\def\draft2{
        \def\@oddfoot{\sl preliminary draft \hfil
        \rm\thepage\hfil\sl\today\quad\militarytime}
        \let\@evenfoot\@oddfoot \overfullrule 3pt
        \let\label=\draftlabel
        \let\marginnote=\draftmarginnote
   \def\@eqnnum{(\theequation)\rlap{\kern\marginparsep\tt\@eqnlabel}%
\global\let\@eqnlabel\@vacuum}  }

\def\preprint{\twocolumn\sloppy\flushbottom\parindent 2em
        \leftmargini 2em\leftmarginv .5em\leftmarginvi .5em
        \oddsidemargin -.5in    \evensidemargin -.5in
        \columnsep .4in \footheight 0pt
        \textwidth 10.in        \topmargin  -.4in
        \headheight 12pt \topskip .4in
        \textheight 6.9in \footskip 0pt
        \def\@oddhead{\thepage\hfil\addtocounter{page}{1}\thepage}
        \let\@evenhead\@oddhead \def\@oddfoot{} \def\@evenfoot{} }

\def\numberbysection{\@addtoreset{equation}{section}
        \def\theequation{\thesection.\arabic{equation}}}

\def\underline#1{\relax\ifmmode\@@underline#1\else
        $\@@underline{\hbox{#1}}$\relax\fi}

\def\titlepage{\@restonecolfalse\if@twocolumn\@restonecoltrue\onecolumn
     \else \newpage \fi \thispagestyle{empty}\c@page\z@
        \def\thefootnote{\fnsymbol{footnote}} }

\def\endtitlepage{\if@restonecol\twocolumn \else \newpage \fi
        \def\thefootnote{\arabic{footnote}}
        \setcounter{footnote}{0}}  %\c@footnote\z@ }

\catcode`@=12 \relax

\def\figcap{\section*{Figure Captions\markboth
        {FIGURECAPTIONS}{FIGURECAPTIONS}}\list
        {Figure \arabic{enumi}:\hfill}{\settowidth\labelwidth{Figure
999:}
        \leftmargin\labelwidth
        \advance\leftmargin\labelsep\usecounter{enumi}}}
 \relax
\def\tablecap{\section*{Table Captions\markboth
        {TABLECAPTIONS}{TABLECAPTIONS}}\list
        {Table \arabic{enumi}:\hfill}{\settowidth\labelwidth{Table
999:}
        \leftmargin\labelwidth
        \advance\leftmargin\labelsep\usecounter{enumi}}}
 \relax
\def\reflist{\section*{References\markboth
        {REFLIST}{REFLIST}}\list
        {[\arabic{enumi}]\hfill}{\settowidth\labelwidth{[999]}
        \leftmargin\labelwidth
        \advance\leftmargin\labelsep\usecounter{enumi}}}
 \relax
\makeatletter
\newcounter{pubctr}
\def\publist{\@ifnextchar[{\@publist}{\@@publist}}
\def\@publist[#1]{\list
        {[\arabic{pubctr}]\hfill}{\settowidth\labelwidth{[999]}
        \leftmargin\labelwidth
        \advance\leftmargin\labelsep
        \@nmbrlisttrue\def\@listctr{pubctr}
        \setcounter{pubctr}{#1}\addtocounter{pubctr}{-1}}}
\def\@@publist{\list
        {[\arabic{pubctr}]\hfill}{\settowidth\labelwidth{[999]}
        \leftmargin\labelwidth
        \advance\leftmargin\labelsep
        \@nmbrlisttrue\def\@listctr{pubctr}}}
 \relax
\makeatother

\def\be{\begin{equation}}
\def\ee{\end{equation}}
\def\ba{\begin{eqnarray}}
\def\ea{\end{eqnarray}}

\def\no{\noindent}

\def\IR{\relax{\rm I\kern-.18em R}}

\def\bse{\begin{small}\begin{equation*}}
\def\ese{\end{equation*}\end{small}}

\begin{document}
%\draft2

%\renewcommand{\theequation}{\arabic{equation}}
%\renewcommand{\theequation}{\thesection.\arabic{equation}}

\renewcommand{\theequation}{\thesection.\arabic{equation}}
\csname @addtoreset\endcsname{equation}{section}

\newcommand{\eqn}[1]{(\ref{#1})}

\begin{titlepage}
\begin{center}
\strut\hfill
\vskip 1.3cm

%\hfill  [hep-th]\\

\vskip .5in

{\Large \bf A classical variant of the vertex algebra $\&$ the auxiliary linear problem}

\vskip 0.5in

{\large {\bf Anastasia Doikou}} \vskip 0.2in

 \vskip 0.02in
{\footnotesize
 Department of Mathematics, Heriot-Watt University,\\
EH14 4AS, Edinburgh, United Kingdom}
\\[2mm]
\noindent
{\footnotesize and}

\vskip 0.02in
{\footnotesize Department of Computer Engineering \& Informatics,\\
 University of Patras, GR-26500 Patras, Greece}
\\[2mm]
\noindent
\vskip .1cm

%\vskip -.15in

{\footnotesize {\tt E-mail: a.doikou@hw.ac.uk}}\\

\end{center}

\vskip 1.0in

\centerline{\bf Abstract}
We propose a classical analogue of the vertex algebra in the context of classical integrable field theories.
We use this fundamental notion to describe the auxiliary function of the linear auxiliary problem as a classical vertex operator. Then using the underlying algebra satisfied by the auxiliary function together with the linear auxiliary problem we identify the local integrals of motion, which by construction are in involution. The time components of the Lax pair are also identified in terms of the classical vertex operators. Systems in the presence of point like defects as well as systems on the semi-infinite line are investigated. Specific examples associated to the classical Yangian and twisted Yangian are also presented.

\no

\vfill

\end{titlepage}
\vfill \eject

%\def\baselinestretch{1.2}
%\baselineskip 10 pt
%\noindent

\tableofcontents

\section{Introduction}

The starting point in the present analysis is the notion of vertex operators at the quantum level and the associated Faddeev-Zamolodchikov algebra \cite{Faddeev-Zamolodchikov}. Inspired by these ideas we propose a similar algebraic formulation to deal with classical integrable field theories on the infinite or semi-infinite line. It is worth noting that such ideas at the classical level were briefly discussed in \cite{Lukyanov}, but a systematic construction of classical vertex operators, the generating function of the local integrals of motion as well as a construction of the time component of the Lax pair in terms of the classical vertex operators was not really demonstrated. We should stress that one of the key points of the present analysis is the identification of the auxiliary function of the auxiliary linear problem as the classical version of the vertex operator. Moreover, particular emphasis in the present investigation is given in integrable systems in the presence of point like defects as well as in the presence of non-trivial boundary conditions.

Recall that vertex operators $\Phi$ at the quantum level satisfy the Faddeev-Zamolodchikov algebra \cite{Faddeev-Zamolodchikov}
\be
S_{12}(\lambda_1 -\lambda_2)\ \Phi_1(\lambda_1)\ \Phi_2(\lambda_2) = \Phi_2(\lambda_2)\ \Phi_1(\lambda_1),\label{quantum-vertex}
\ee
where $S$ is the physical scattering matrix $S \in \mbox{End}({\mathbb C}^{\cal N} \otimes {\mathbb C}^{\cal N})$, solution of the Yang-Baxter equation, and $\Phi$ is the quantum vertex operator which is an ${\cal N}$-dimensional vector with elements being realizations of the underlying affine quantum algebra. Here we are going to formulate a classical analogue of the latter quantum objects $\Phi$ as well as the algebra (\ref{quantum-vertex}). This will be basically achieved by means of certain vectors --tensor products of which are eigenvectors of the classical $r$-matrix-- suitably implemented at the ends of the system.

The paper is organized as follows: in the next section we briefly review the auxiliary linear problem and also recall the notion of the classical monodromy matrix together with the corresponding classical quadratic algebra. Then based on these ideas we introduce the classical analogue of the vertex operator, which coincides essentially with the auxiliary function of the auxiliary linear problem, which in turn satisfies a classical version of the vertex algebra (\ref{quantum-vertex}). We then introduce the generating function of the integrals of motion in terms of classical vertex operators. The extracted charges are in involution by construction due to the fact that the classical vertex operators satisfy a quadratic vertex algebra. Based on the classical vertex algebra and the Hamiltonian equations of motion we derive the time components of the Lax pairs. This setting is generalized for systems in the presence if point like defects as well as for systems on the half line.
In section 3 we exemplify our construction using as a paradigm a whole class of systems associated to the classical Yangian. Explicit expressions of the generating functions of the integrals of motion as well as the time components of the Lax pairs are provided.

\section{The general setting}

The starting point in our analysis will be the auxiliary linear problem, consisting of the Lax pair ${\mathbb U},\ {\mathbb V}$ and the pair of differential equations (see \cite{FT} and references therein):
\ba
&& {\partial \Psi \over \partial x } = {\mathbb U}(x, t;\lambda)\ \Psi \cr
&& {\partial \Psi \over \partial t } = {\mathbb V}(x, t;\lambda)\ \Psi. \label{auxiliary}
\ea
${\mathbb U},\ {\mathbb V}$ are in general ${\cal N} \times {\cal N}$ matrices with entries being functions of $x,\ t$ and depending also on a spectral parameter.
The main proposition of the present investigation is that the auxiliary vector functions $\Psi$ are essentially classical variants of the vertex operators, and they satisfy a classical version of the vertex (Faddeev-Zamolodvhikov) algebra (\ref{quantum-vertex}).

Let us recall the classical monodromy matrix, which will be an essential object in our formulation
\be
T(x,y,t; \lambda) = {\cal P} \exp \Big \{ \int_x^y dx'\ {\mathbb U}(x', t;\lambda)\Big \},
\ee
is a solution of the first of the equations (\ref{auxiliary}). Moreover, the monodromy matrix satisfies the quadratic classical algebra\footnote{Note that throughout the text the dependence on $x,\ t$ is implied even if it is not explicitly stated.}
\be
\Big \{T_1(\lambda_1),\ T_2(\lambda_2) \Big \} = \Big [r_{12}(\lambda_1 -\lambda_2),\ T_1(\lambda_1)\  T_2(\lambda_2) \Big ], \label{algebra1}
\ee
where as usual we introduce the notation: $T_1 = T \otimes {\mathbb I},\ T_2 = {\mathbb I} \otimes T$ and so on. The classical $r$-matrix satisfies the classical Yang-Baxter equation \cite{sts}
\be
\Big [r_{12}(\lambda_1 -\lambda_2),\ r_{13}(\lambda_1) +r_{23}(\lambda_2) \Big ] + \Big [r_{13}(\lambda_1),\ r_{23}(\lambda_2) \Big ] =0. \label{clasYBE}
\ee

We shall be mainly interested henceforth in integrable models on the full line or on the semi-infinite line. We shall begin our derivation with integrable systems on the full line considering the interval $[-L,\ L]$. We define the following basic objects
\ba
&& \Psi(x, -L,t; \lambda) = T(x, -L,t; \lambda)\ V \cr
&& \Psi^*(L,x, t; \lambda) = V^*\ T(L,x,t ;\lambda), \label{definitions}
\ea
where $V$, $V^*$ are ${\cal N}$ row and column vectors respectively such that:
\be
V_1^*\ V_2^*\ r_{12}(\lambda) = f(\lambda)\ V_1^*\ V_2^*, ~~~~~r_{12}(\lambda)\ V_1\ V_2 = f(\lambda)\ V_1\ V_2 \label{const1}
\ee
$f(\lambda)$ a function of the spectral parameter depending on the choice of the $r$-matrix.

Then it is straightforward to show using the algebraic relation (\ref{algebra1}) and the definitions (\ref{definitions}), (\ref{const1}):
\ba
&& \Big \{ \Psi_1(\lambda_1), \Psi_2(\lambda_2) \Big \} =  {\mathfrak r}_{12}(\lambda_1 - \lambda_2)\ \Psi_1(\lambda_1)\ \Psi_2(\lambda_2) \cr
&& \Big \{ \Psi_1^*(\lambda_1), \ \Psi_2^*(\lambda_2)\Big \} =  - \Psi_1^*(\lambda_1)\  \Psi_2^*(\lambda_2)\  {\mathfrak r_{12}}(\lambda_1 -\lambda_2) \label{vertexI}
\ea
(see also similar relations at the classical level in \cite{Lukyanov}), we define the ``shifted'' classical $r$-matrix:
\be
{\mathfrak r}(\lambda) = r(\lambda) - f(\lambda)
\ee
Note that we consdider here a quite generic classical $r$-matrix: $r \in \mbox{End}({\mathbb C}^{{\cal N}} \otimes {\mathbb C}^{{\cal N}})$. In the last section we are going to restrict our attention to the $\mathfrak{gl}_{\cal N}$ classical Yangian.

Keeping in mind the latter relations we can introduce the fundamental quantity:
\be
{\mathbb T}(L, -L, t; \lambda) = \Psi^*(L, 0,t; \lambda)\ \Psi(0, -L, t; \lambda). \label{funda1}
\ee
The latter (\ref{funda1}) provides the charges in involution of the system under consideration, indeed it may shown via (\ref{vertexI})
that
\be
\Big \{{\mathbb T}_1(\lambda_2),\ {\mathbb T}_2(\lambda_2) \Big \} = 0. \label{commut1}
\ee
Expansion of the $\ln({\mathbb T})$ clearly provides the local charges in involution (local integrals of motion). Later in the text we are going to consider a class of integrable models associated to classical Yangians \cite{yang}, with typical examples e.g the generalized NLS and the Landau-Lifshitz models.

Employing the formulation described above we may extract the hierarchy of charges as in the inverse scattering process. The main advantage here is the fact that the auxiliary function may be thought of as a classical vertex operator satisfying the classical algebras (\ref{vertexI}), and thus involution of the charges is by construction guaranteed. Moreover, due to the fact that the classical vertex operators are immediately associated to the monodromy matrix facilitates technically the derivation of the desired quantities. The time depended equation of the pair (\ref{auxiliary}) is not really used as opposed to the inverse scattering methodology where the time dependent equation is exploited in order to prove the conservation of the extracted local charges. In fact, the time components of the Lax pair associated to each integral of motion can be also derived in terms the classical vertex operators.

Indeed, we shall derive below the ${\mathbb V}$-operators in terms of the classical vertex operators exploiting the underlying algebra. Recall the fundamental Poisson commutator (see e.g. \cite{FT})
\ba
&&\Big \{ T_1(A, B,t;\lambda),\ {\mathbb U}_2(x,t;\mu) \Big \} = {\partial {\mathbb M}_{12}(x, t;\lambda, \mu) \over \partial x} + \Big [ {\mathbb M}_{12}(x, t;\lambda, \mu),\ {\mathbb U}_2(x,t;\mu)  \Big ],
\cr && x \in [A,\ B],
\ea
where we define
\be
{\mathbb M}_{12}(x,t; \lambda, \mu) = T_1(A, x,t ;\lambda)\ r_{12}(\lambda -\mu)\ T_1(x, B,t; \mu)
\ee

Recall also the definitions for $\Psi$ and $\Psi^*$ and formulate:
\be
\Big \{\Psi_1^*(L, 0,t; \lambda)\ \Psi_1(0, -L,t; \lambda),\ {\mathbb U}_2(x,t;\mu) \Big \} = {\partial {\mathbb N}_{2}(x,t;\lambda,\mu) \over
\partial x} + \Big [ {\mathbb N}_{2}(x,t;\lambda, \mu),\  {\mathbb U}_2(x,t;\mu) \Big ]
\ee
where we define
\be
{\mathbb N}_{2}(x,t; \lambda, \mu) = \Psi_1^*(L,x,t ;\lambda)\ r_{12}(\lambda -\mu)\ \Psi_1(x, -L,t;\lambda)
\ee

We wish to consider $\ln({\mathbb T})$  as the generating function of the local integrals of motion
as well as the zero curvature condition arising as a compatibility condition of (\ref{auxiliary})
\be
{\mathbb V}'(x,t; \lambda) - \dot {\mathbb U}(x, t; \lambda) + \Big [{\mathbb V}(x,t; \lambda),\ {\mathbb U}(x,t; \lambda) \Big ] =0.
\ee
the ``prime'' denotes derivation with respect to $x$, and the ``dot'' denotes derivation with respect to $t$. Then
\be
\Big \{\ln\Big (\Psi_1^*(L, 0,t; \lambda)\ \Psi_1(0, -L,t; \lambda)\Big ), \ {\mathbb U}_2(x,t;\mu) \Big \} = {\partial {\mathbb V}_{2}(x,t;\lambda,\mu) \over
 \partial x} + \Big [ {\mathbb V}_{2}(x,t;\lambda, \mu),\  {\mathbb U}_2(x,t;\mu) \Big ]
\ee
where
\be
{\mathbb V}_{2}(x,t;\lambda, \mu) = {\mathbb T}(\lambda)^{-1}\ \Psi^*_1(L,x,t ;\lambda)\ r_{12}(\lambda -\mu)\ \Psi_1(x, -L,t;\lambda) \label{vv1}
\ee
Expansion of ${\mathbb V}$ in powers of $\lambda^{-1}$ provides the hierarchy of ${\mathbb V}$-operators associated to each integral of motion.

\subsection{Defects}

The approach described above may be easily generalized in the presence of local integrable defects (see also e.g. \cite{haku}--\cite{caudrelier} for recent results in this context).
In fact, the relevant object in this case, i.e. the generating function of the integrals of motion would be
\be
{\mathbb D}(\lambda) = \Psi^{*+}(L, x_0,t;\lambda)\ {\mathbb L}(x_0,t; \lambda)\ \Psi^-(x_0, -L,t;\lambda) \label{gener2}
\ee
the plus and minus superscripts in the expression above refer to the left right bulk theories, the defect matrix ${\mathbb L}$ is of the generic form
\be
{\mathbb L}(\lambda) = \sum_{i,j =1}^{{\cal N}} {\mathbb L}_{ij}(\lambda)\ e_{ij}, \label{defectb}
\ee
$e_{ij}$ are in general ${\cal N} \times {\cal N}$ matrices such that: $(e_{ij})_{kl} = \delta_{ik}\ \delta_{jl}$.
${\mathbb L}$ satisfies the quadratic algebra
\be
\Big \{{\mathbb L}_1(\lambda_2),\ {\mathbb L}_2(\lambda_2) \Big \} = \Big [{\mathfrak r}_{12}(\lambda_1 - \lambda_2),\ {\mathbb L}_1(\lambda_2)\ {\mathbb L}_2(\lambda_2) \Big ]. \label{defect}
\ee
Exploiting the algebraic relations between the vertices (\ref{vertexI}) as well as the ones regarding the defect matrix (\ref{defect}) we conclude as expected that ${\mathbb D}$ provides the charges in involution,
\be
\Big \{{\mathbb D}(\lambda_1),\ {\mathbb D}(\lambda_2) \Big \} = 0, \label{commut2}
\ee
i.e. the involution of the charges derived via the expansion of the generating function ${\mathbb D}$ is guaranteed by construction.

We shall derive below the ${\mathbb V}$-operator associated to the defect point computed from the left and from the right in terms of the classical vertex operators. The ${\mathbb V}$-operators at $x \neq x_0$ coincide with the bulk expressions for the left and right theories, and we do not rederive them here for brevity.
To derive the ${\mathbb V}$-operator on the defect point we recall the zero curvature condition at $x = x_0$ \cite{avan-doikou}
\be
{\partial {\mathbb L}(x_0,t;\lambda) \over \partial t} = \tilde {\mathbb V}^+(x_0,t; \lambda)\ {\mathbb L}(x_0,t;\lambda) - {\mathbb L}(x_0,t;\lambda)\ \tilde {\mathbb V}^-(x_0,t;\lambda)
\ee
Then using the algebraic relations (\ref{defect}) we may formulate on the defect point:
\ba
&&\Big \{\ln\Big (\Psi_1^{*+}(L,x_0,t;\lambda)\ {\mathbb L}_1(x_0,t; \lambda)\ \Psi_1^-(x_0, -L,t;\lambda)\Big ),\ {\mathbb L}_2(x_0,t;\mu) \Big \} \cr
&& = \tilde {\mathbb V}^+_2(x_0,t;\lambda,\mu)\ {\mathbb L}_2(x_0,t; \mu)- {\mathbb L}_2(x_0,t; \mu)\ \tilde{\mathbb V}_2^-(x_0,t; \lambda, \mu)
\ea
where $\tilde {\mathbb V}^{\pm}$ are the time components of the Lax pair on the defect point computed from the left and right:
\ba
&& \tilde {\mathbb V}_2^{+}(x_0,t; \lambda, \mu)= {\mathbb D}^{-1}(x_0,t;\lambda)\ \Psi_1^{*+}(L, x_0,t; \lambda)\ r_{12}(\lambda - \mu)\ {\mathbb L}_1(x_0,t;\lambda)\ \Psi_1^-(x_0, -L,t; \lambda) \cr
&& \tilde {\mathbb V}_2^{-}(x_0,t; \lambda, \mu)= {\mathbb D}^{-1}(x_0,t;\lambda)\ \Psi_1^{*+}(L, x_0,t; \lambda)\ {\mathbb L}_1(x_0,t;\lambda)\ r_{12}(\lambda - \mu)\ \Psi_1^-(x_0, -L,t; \lambda). \cr &&\label{vv2}
\ea
Let us recall, as has been discussed in detail in \cite{avan-doikou}, that analyticity conditions imposed on the ${\mathbb V}$-operators around the defect point, i.e.
\be
\tilde {\mathbb V}^{\pm}(x_0) \to {\mathbb V}^{\pm}(x_0^{\pm})
\ee
lead to the necessary sewing conditions on the defect point. These are conditions that involve the left and right fields and their derivatives and the defect degrees of freedom. Due to these conditions one observes ``jumps'' of the fields and the derivative across the defect point.

\subsection{Integrable systems on the half line}

We shall consider in this section the generalization of the aforementioned algebraic setting for integrable systems on the half line. Before we proceed with the mathematical setting in this case it will be instructive to recall the two distinct types of boundary conditions appearing in integrable theories associated to higher rank algebra i.e. $\mathfrak{gl}_{{\cal N}}$. In general, depending on the choice of boundary conditions the bulk physical behavior may be accordingly affected. In the frame of affine Toda field theories (ATFTs) the boundary conditions introduced in \cite{durham} are related to the classical twisted Yangian \cite{twisted-yangian}, and force a soliton to reflect to an anti-soliton, hence the relevant appellation soliton non-preserving (SNP) boundary conditions. Naturally another possibility exists, that is boundary conditions that lead to the reflection of a soliton to itself. These boundary conditions are associated to classical reflection algebras \cite{sklyanin} are known as soliton preserving (SP), and have been extensively investigated in the context of integrable quantum spin chains (see e.g. \cite{doikou-nepo} and references therein). Although SP boundary conditions are the obvious ones in the framework of integrable lattice models they remained puzzling in the context of ATFT’s until their thorough analysis in \cite{doikou-atft}. On the other hand SNP boundary conditions were investigated through the Bethe
ansatz formulation for the first time in \cite{doikou-twisted}, whereas higher rank generalizations considered in \cite{annecy}.

Let us now describe the algebraic setting associated to integrable systems on the semi-infinite line. Recall that the modified monodromy matrix for a system on the half line is given by \cite{sklyanin}
\be
{\cal T}(\lambda) = \hat T(\lambda)\ K(\lambda)\ T(\lambda) \label{1}
\ee
where we define $\hat T$ in the case one considers the twisted Yangian as:
\be
\hat T(\lambda) = U\ T^t(-\lambda)\ U \label{2}
\ee
$U$ can be a diagonal or anti-diagonal matrix with particular form depending on the choice of the classical $r$-matrix. We shall later focus on the classical Yangian and consider $U = \mbox{antidiag}(1, 1, \ldots, 1)$.
In the case one considers the reflection algebra $\hat T$ is defined as \cite{sklyanin}
\be
\hat T(\lambda) = T^{-1}(-\lambda).
\ee

We shall focus henceforth on the twisted Yangian case mainly because this case has not been really explored in the context of classical continuum field theories such as the vector NLS models or the generalized Landau-Lifshitz model. In this case due to (\ref{1}), (\ref{2}) it may by easily shown that
${\cal T}$ satisfies
\ba
\Big \{{\cal T}_1(\lambda),\ {\cal T}_2(\mu) \Big \} &=& r_{21}(\lambda_1 -\lambda_2)\ {\cal T}_1(\lambda_2)\  {\cal T}_2(\lambda_2) - {\cal T}_1(\lambda_1)\ {\cal T}_2(\lambda_2)\ r_{12}(\lambda_1 -\lambda_2) \cr &+& {\cal T}_1(\lambda_1)\ \bar r_{12}(\lambda_1 +\lambda_2)\ {\cal T}_2(\lambda_2) - {\cal T}_2(\lambda_1)\ \bar r_{12}(\lambda_1 + \lambda_2)\ {\cal T}_1(\lambda_1) \label{twisted}
\ea
where we define in the SNP case (classical twisted Yangian)
\be
\bar r(\lambda ) = U_1\ r^{t_2}_{12}(\lambda)\ U_1,
\ee
$K$ is a $c$-number solutions of the twisted Yangian (\ref{twisted}),
\ba
&& r_{21}(\lambda_1 -\lambda_2)\ K_1(\lambda_2)\ K_2(\lambda_2) - K_1(\lambda_1)\ K_2(\lambda_2)\ r_{12}(\lambda_1 -\lambda_2) \cr
&& +K_1(\lambda_1)\ \bar r_{12}(\lambda_1 +\lambda_2)\ K_2(\lambda_2) - K_2(\lambda_1)\ \bar r_{12}(\lambda_1 + \lambda_2)\ K_1(\lambda_1) =0 \label{c-number}
\ea
where here notice that in the $\mathfrak{sl}_2$ case the twisted Yangian and reflection algebras coincide. Henceforth we shall adopt the twisted Yangian for our computations, extra attention will be given to the the $\mathfrak{gl}_{\cal N}$, ${\cal N} > 2$ case in this context. In the case associated to the reflection algebra (SP boundary conditions) $\bar r_{12} = r_{12}$. Although this case is equally interesting it involves various technical intricacies and will be left for future investigations.

In addition to the classical vertex $\Psi$ (\ref{definitions}) we also introduce
\be
\hat \Psi(\lambda) = \hat V\ \hat T(\lambda). \label{definition2}
\ee
Moreover, the following relations are required:
\ba
&& \hat V_1\ \bar r_{12}(\lambda)\ V_2 = f(\lambda)\ \hat V_1\ V_2, \cr
&& \hat V_2\ \bar r_{12}(\lambda)\ V_1 = f(\lambda)\ \hat V_2\ V_1 \cr
&& \hat V_1\ \hat V_2\ r_{12}(\lambda) = f(\lambda)\ \hat V_1\ \hat V_2
\ea
Then in addition to (\ref{vertexI}) the following set of algebraic relations are also satisfied:
\ba
&& \Big \{\hat \Psi_1(\lambda_1),\ \Psi_2(\lambda_2)\Big \} = \hat \Psi_1(\lambda_1)\ \bar {\mathfrak r}_{21}(\lambda_1 +\lambda_2)\ \Psi_2(\lambda_2) \cr
&&\Big \{\Psi_1(\lambda_1),\ \hat \Psi_2(\lambda_2)\Big \} = \Psi_1(\lambda_1)\ \bar {\mathfrak r}_{12}(\lambda_1 +\lambda_2)\ \hat \Psi_2(\lambda_2) \cr
&&\Big \{\hat \Psi_1(\lambda_1),\ \hat \Psi_2(\lambda_2)\Big \} = \hat \Psi_1(\lambda_1)\ \hat \Psi_2(\lambda_2)\ {\mathfrak r}_{21}(\lambda_1 -\lambda_2) \label{vertexII}
\ea
where the shifted $\bar r$-matrix is defined as
\be
\bar {\mathfrak r} = \bar r(\lambda) - f(\lambda). \ee

Define the generating function of the integrals of motion as:
\be
{\mathfrak T}(\lambda) = \hat \Psi(0, -L,t; \lambda)\ K(\lambda)\ \Psi(0, -L,t; \lambda). \label{gener3}
\ee
Indeed taking into account the relations (\ref{vertexI}), (\ref{vertexII}) and (\ref{c-number}) it is shown that
\be
\Big \{ {\mathfrak T}(\lambda_1),\ {\mathfrak T}(\lambda_2)\Big \} =0.
\ee

We may now derive the ${\mathbb V}$-operator on the boundary point $x=0$ (see also  \cite{avan-doikou-boundary} for relevant results). The main aim is to formulate the following Poisson commutator taking into account the algebraic relations as well as the form of the zero curvature condition at $x=0$:
\ba
&& \Big \{\ln\Big (\hat \Psi_1(0, -L,t; \lambda)\ K_1(\lambda)\ \Psi_1(0, -L,t; \lambda)\Big ),\ {\mathbb U}_2(0,t; \mu) \Big \} =\cr
&&  {\partial \tilde {\mathbb V}_{2}(x,t;\lambda, \mu) \over \partial x} + \Big [ \tilde {\mathbb V}_{2}(0,t;\lambda, \mu),\ {\mathbb U}_2(0,t;\mu)\Big ]
\ea
where we define the boundary $\tilde{\mathbb V}$-operator at $x = 0$
\be
\tilde {\mathbb  V}_{2}(\lambda, \mu) = {\mathfrak T}^{-1}(\lambda) \Big (\hat \Psi_1(\lambda)\ K_1(\lambda)\ r_{12}(\lambda-\mu)\ \Psi_1(\lambda) + \hat \Psi_1(\lambda)\ \bar r_{12}(\lambda+\mu)\ K_1(\lambda)\ \Psi_1(\lambda) \Big ).
\ee
Continuity conditions at $x =0$
\be
\tilde {\mathbb V}(0) \to {\mathbb V}(0^+)
\ee
lead to suitable boundary conditions on the fields and their derivatives (see detailed discussion on \cite{avan-doikou-boundary}).

Having introduced all the necessary algebraic objects for classical integrable systems on the full and half line as well as in the presence of point-like defects we can use a class of models associated to the classical (twisted) Yangian $r$-matrix as a paradigm. Particular emphasis will be given on such models with boundary conditions related to higher rank algebras.

\section{Paradigm: the classical Yangian \& twisted Yangian}

Having at our disposal the general algebraic setting in treating integrable systems via the classical version of the vertex algebra we may now consider as a paradigm a whole class of integrable models
associated to the classical $\mathfrak{gl}_{\cal N}$ Yangian with the classical $r$-matrix being \cite{yang}
\be
r(\lambda)={\kappa \over \lambda} \sum_{i, j=1}^{{\cal N}} e_{ij} \otimes e_{ji} \label{yang1}
\ee
in this case $f(\lambda) = {\kappa \over \lambda}$.

We shall first consider the model on the full line and demonstrate the simplicity of obtaining the integrals of motion. Involution is guaranteed by construction via commutation relations (\ref{commut1}). As already mentioned the fact that the classical vertex operator can be immediately obtained from the monodromy matrix provides an elegant way of expressing the classical vertex using the standard decomposition of the monodromy matrix \cite{FT}
\ba
&& T(0, -L,t;\lambda) = (1 + W(0,t;\lambda))\ e^{Z(0, -L,t;\lambda)}\ (1 + F(-L,t;\lambda)), \cr
&& T^{*}(L, 0,t;\lambda) = (1 + W(L,t;\lambda))\ e^{Z(L,0,t;\lambda)}(1+F(0,t;\lambda)) \label{decompositionA}
\ea
where
\be
Z= \sum_j Z_j\ e_{jj}, ~~~~1+F = (1+W)^{-1},~~~~ W = \sum_{i\neq j} W_{ij\ }e_{ij}, ~~~~ F= \sum_{i, j} F_{ij}\ e_{ij}
\ee
$W$ and $Z$ may be identified using the fact the monodromy matrix satisfies the first of the equations (\ref{auxiliary}), and they of course depend on the choice of the model, i.e. the ${\mathbb U}$-operator \cite{FT}.
More precisely, substitution of the decomposition of the type (\ref{decompositionA}) for $T(x, y; \lambda)$ in the first of the pair of equations (\ref{auxiliary}) leads to typical Ricatti equations for $W$ \cite{FT}. Solving these one can identify $W$ and $Z$.

We assume Schwartz boundary conditions at the end points $x =\pm L$, i.e. the fields and their derivatives disappear so that $W(\pm L) =0$. Introduce also the ${\cal N}$ dimensional column vector ${\mathrm u}_k$ with unit on the $k^{th}$ position and zeros elsewhere, also $\hat  {\mathrm u}_k =  {\mathrm u}_k^t$.
Also we choose to consider henceforth
\be
V = {\mathrm u}_{\cal N}, ~~~~~~V^* = \hat  {\mathrm u}_{\cal N}.
\ee
Then the fundamental quantities $\Psi,\ \Psi^*$ with the use of the identities:
\be
e_{ij}\ e_{kl} = \delta_{jk}\ \delta_{il}, ~~~~e_{ij}\ {\mathrm u}_k =\delta_{jk}\ {\mathrm u}_i, ~~~~~\hat {\mathrm u}_k\ e_{ij} = \delta_{ik} \hat {\mathrm u}_j,~~~~\hat {\mathrm u}_i\ {\mathrm u}_j = \delta_{ij}
\ee
become
\ba
&& \Psi(0,-L,t;\lambda) = e^{Z_{\cal N}(0, -L,t; \lambda)} \Big ( {\mathrm u}_{\cal N} + \sum_{j\neq {\cal N}} W_{j{\cal N}}(0,t; \lambda)\ {\mathrm u}_j\Big ) \cr
&& \Psi^*(L, 0,t;\lambda) = e^{Z_{\cal N}(L, 0,t; \lambda)} \Big ( \hat {\mathrm u}_{\cal N} + \sum_j F_{{\cal N}j}(0,t; \lambda)\ \hat {\mathrm u}_j\Big ). \label{psi}
\ea
Note that in general we could have used the  standard inverse methodology to identify the the rations ${\Psi^{(j)} \over \Psi^{(1)}},\ {\psi^{*(j)} \over \Psi^{*(1)}}$, where obviously and equivalently to (\ref{psi})
\be
\Psi = \sum_{j=1}^{{\cal N}} \Psi^{(j)}\ {\mathrm u}_j, ~~~~~\Psi^* = \sum_{j=1}^{{\cal N}} \Psi^{*(j)}\ \hat {\mathrm u}_j.
\ee
Indeed, recalling that $\Psi$ satisfies the first of the equations (\ref{auxiliary}) and expressing the ${\mathbb U}$ operator as
\be
{\mathbb U}(x,t;\lambda) = \sum_{i, j} {\mathbb U}_{ij}(x,t;\lambda)\ e_{ij},
\ee
one obtains
\be
{\partial \over \partial x}\Big [e^{Z_{\cal N}(x,-L,t)}{\mathrm u}_{\cal N} + e^{Z_{\cal N}(x, -L,t)} \sum_{j\neq {\cal N}}W_{j{\cal N}}(x,t) {\mathrm u}_j\Big ] = \sum_{i}
{\mathbb U}_{i{\cal N}}(x,t){\mathrm u}_{\cal N} + \sum_{j\neq {\cal N}, i}{\mathbb U}_{ij}(x,t) W_{j{\cal N}}(x,t) {\mathrm u}_i
\ee
which leads to the typical Ricatti type equations for $W_{j{\cal N}}$ already mentioned above
\ba
&& {\partial W_{k {\cal N}}(x,t) \over \partial x}+  \Big ( {\mathbb U}_{{\cal N} {\cal N}}(x,t) + \sum_{j\neq1} {\mathbb U}_{{\cal N} j}(x,t)W_{j {\cal N}}(x,t)\Big ) W_{k {\cal N}}(x)= \cr
&&  {\mathbb U}_{k {\cal N}}(x,t)+ \sum_{j\neq {\cal N}} {\mathbb U}_{kj}(x,t) W_{j {\cal N}}(x,t) \cr
&& {\partial Z_{\cal N}(x,t) \over \partial x} = {\mathbb U}_{{\cal N} {\cal N}}(x,t) + \sum_{j \neq {\cal N}} {\mathbb U}_{{\cal N}j}(x,t) W_{j{\cal N}}(x,t),
\ea
solution of the equations above leads to the identification of $Z_{\cal N}$, $~W_{j {\cal N}}$.

The advantage here is the use of the decomposition of the monodromy matrix (\ref{decompositionA}), which provides the simple formula below for the generating function of the integrals of motion:
\be
\ln({\mathbb T}(L, -L, t;\lambda)) = Z_{\cal N}(L, -L,t;\lambda)
\ee
with apparent analyticity conditions at $x=0$, this will become more transparent when discussing such models in the presence of point-like defects later in the text.
It will be also instructive to derive the ${\mathbb V}$-operator (\ref{vv1}). Taking into account (\ref{vv1}), (\ref{yang1}), (\ref{psi}) we conclude that
\be
{\mathbb V}(x,t; \lambda, \mu) = {\kappa \over \lambda- \mu} \Big (e_{{\cal N} {\cal N}}+ \sum_{j}  F_{{\cal N} j}(\lambda)\ e_{{\cal N} j} + \sum_{j \neq {\cal N}} W_{j {\cal N}}(\lambda)\ e_{j {\cal N}} + \sum_{i\, j\neq {\cal N}} F_{{\cal N} i}(\lambda)\ W_{j {\cal N}}(\lambda)\ e_{ji}\Big ),
\ee
which coincides with the generic expression for ${\mathbb V}$ in the context of classical Yangian \cite{FT, avan-doikou}. Substitution of the $W_{ij}$ and $Z_{\cal N}$ quantities on the expressions for the generating function and the ${\mathbb V}$-operator provides explicit expressions at each order for a particular model under study (see e.g. \cite{FT, avan-doikou} for explicit expressions).

\subsection{Defects}

The main aim in this subsection is to identify explicit expressions of the generating function and hence the local integrals of motion as well the time component of the Lax pair using the classical vertex operators in the presence of a local defect for the classical Yangian case.

Recall the expression for the generating function (\ref{gener2}), it is then straightforward to rewrite it via (\ref{defectb}), (\ref{psi}),
as
\be
\ln ({\mathbb D}(\lambda)) = Z_{{\cal N}}^+(L, x_0,t; \lambda) + Z_{{\cal N}}^-(x_0, -L,t; \lambda) + \ln {\mathrm X}(x_0,t;\lambda)
\ee
the plus and minus superscripts in the expression above refer to the left right bulk theories, and define
\be
{\mathrm X}(x_0,t;\lambda) ={\mathbb L}_{{\cal N}{\cal N}}(\lambda)+ \sum_{j}  F^+_{{\cal N}j}(\lambda)\ {\mathbb L}_{j{\cal N}}(\lambda) + \sum_{j\neq {\cal N}}{\mathbb L}_{{\cal N}j}(\lambda)\ W^-_{j{\cal N}}(\lambda) + \sum_{j\neq {\cal N}} F^+_{{\cal N}i}(\lambda)\ {\mathbb L}_{ij}(\lambda)\ W^-_{j{\cal N}}(\lambda)
\ee

Similarly the time components of the Lax pair on the defect point (\ref{vv2}), take the explicit expressions after substituting (\ref{yang1}), (\ref{defectb}), (\ref{psi}) in (\ref{vv2})
\ba\tilde{\mathbb V}^+(\lambda, \mu) = {\kappa\ {\mathrm X}^{-1} \over \lambda -\mu} \Big (\sum_j {\mathbb L}_{j{\cal N}} e_{j{\cal N}} + \sum_{i \neq 1, j} W^-_{i{\cal N}}{\mathbb L}_{ji} e_{j{\cal N}} + \sum_{i, j} F^+_{{\cal N}j} {\mathbb L}_{i{\cal N}} e_{ij} +\sum_{i\neq {\cal N}, j, l} F^+_{{\cal N}j} W^-_{i{\cal N}} {\mathbb L}_{li} e_{lj} \Big ) \nonumber
\ea
\ba
\tilde {\mathbb V}^-(\lambda, \mu) = {\kappa\ {\mathrm X}^{-1} \over \lambda -\mu} \Big (\sum_j {\mathbb L}_{{\cal N}j} e_{{\cal N}j} + \sum_{i \neq {\cal N}, j} W^-_{i{\cal N}}{\mathbb L}_{{\cal N}j} e_{ij} + \sum_{i, j} F^+_{j{\cal N}} {\mathbb L}_{ji} e_{{\cal N}i} +\sum_{i\neq {\cal N}, j, k} F^+_{{\cal N}j} W^-_{i{\cal N}} {\mathbb L}_{jk} e_{ik} \Big ) \nonumber
\ea
The expressions of the generating function and the defect ${\mathbb V}$-operators coincide as expected with the ones of earlier works on models with underlying classical Yangian \cite{avan-doikou}. Explicit expressions associated e.g. to the NLS model and $\sigma$ models in the presence of  defects can be found in \cite{avan-doikou}, and they are not repeated here for brevity. The interested reader however is referred to \cite{avan-doikou} for more details.

\subsection{The system on the half line}

We shall now derive explicit expressions for any system associated to the classical twisted Yangian such as the vector NLS model or the isotropic Landau-Lifshitz model or principal chiral models. Note that a systematic study at the level of classical field theories is still missing although there are some preliminary results in \cite{doikou-fioravanti-ravanini}, but mainly focused on the discrete analogues of these models. This is in fact the first time that this problem is directly addressed at the level of continuum classical systems.

Let us recall the generating function of the integrals of motion in this case given in (\ref{gener3}), also in addition to the explicit expression for $\Psi$ (\ref{psi}) one needs the expression for $\hat \Psi$. Indeed choosing to consider $\hat V = \hat {\mathrm u}_{{\cal N}}$ we obtain
\be
\hat \Psi(0, -L,t; \lambda) = e^{\hat Z_{\cal N}(0, -L,t; \lambda)} \Big (\hat {\mathrm u}_1 + \sum_{j\neq 1} \hat W_{\bar j {\cal N}}(0,t; \lambda)\hat {\mathrm u}_j \Big )\label{hatpsi}
\ee
where: $\hat Z_{\cal N}(\lambda) = Z_{\cal N}(-\lambda)$, $~\hat W_{ij}(\lambda) = W_{ij}(-\lambda)$, and we also define the conjugate index $\bar j = {\cal N} - j+1$. Also, the $c$-number $K$-matrix can be generically expressed as: $K = \sum_{i, j} K_{ij} e_{ij}$ then the generating function becomes
\ba
&& {\mathfrak T}(\lambda) = e^{Z_{\cal N}(\lambda) + \hat Z_{\cal N}(\lambda)}\ {\mathrm Y}(\lambda)\cr
&& {\mathrm Y}(\lambda)= K_{1 {\cal N}} + \sum_{j \neq {\cal N}} K_{1 j} W_{j{\cal N}} + \sum_{j \neq {\cal N}}K_{j {\cal N}} \hat W_{\bar j {\cal N}} + \sum_{j \neq 1, i \neq {\cal N}} \hat W_{\bar j {\cal N}}K_{ji} W_{i {\cal N}}. \label{gen1}
\ea

Due to the generic expression for the boundary ${\mathbb V}$-operator (\ref{psi}), (\ref{hatpsi}) and
\ba
\tilde {\mathbb V}(0,t;\lambda, \mu) &=&  {\kappa\ {\mathrm Y}^{-1}\over \lambda - \mu} \Big [\sum_l \Big ( K_{1 l } + \sum_{i\neq 1} K_{i {\cal N}} \hat W_{\bar i {\cal N}}\Big ) e_{{\cal N}l} + \sum_{i\neq {\cal N}, l}\Big (K_{1 l } W_{i{\cal N}}  + \sum_{j\neq 1} K_{jl} \hat W_{\bar j {\cal N}} W_{i {\cal N}}  \Big ) e_{il}  \Big ]\cr
&-& {\kappa\ {\mathrm Y}^{-1}\over \lambda + \mu}  \Big [\sum_l \Big ( K_{\bar l {\cal N}} + \sum_{j\neq {\cal N}} K_{\bar l j} W_{ j {\cal N}}\Big ) e_{{\cal N}l} + \sum_{j\neq 1, l}\Big (K_{\bar l {\cal N}} \hat W_{\bar j {\cal N}}  + \sum_{i \neq {\cal N}} K_{\bar l i} \hat W_{\bar j {\cal N}} W_{i {\cal N}} \Big )e_{\bar jl}  \Big ] \cr
&& \label{gen2}
\ea

It is worth presenting here, in addition to the generic expressions (\ref{gen1}), (\ref{gen2}), specific conserved quantities associated to the vector NLS model. In \cite{doikou-fioravanti-ravanini} results on the boundary vector NLS model, but using its discrete counterpart and then taking a suitable continuum limit. Let us now focus on the vector NLS model; the corresponding ${\mathbb U}$-operator
\be
{\mathbb U}(\lambda) ={\mathbb U}_0 + \lambda {\mathbb U}_1
\ee
where ${\mathbb U}_i$ are ${\cal N} \times {\cal N}$ matrices
\ba
{\mathbb U}_1 &=& {1\over 2i} \Big ( \sum_{j=1}^{{\cal N}-1} e_{jj} - e_{{\cal N} {\cal N}}\Big ) \cr
{\mathbb U}_0 &=& \sqrt{\kappa} \sum_{j=1}^{{\cal N}-1} \Big ( \bar \psi_j e_{j{\cal N}} + \psi_j e_{{\cal N} j} \Big )
\ea

Then from the Riccati equations one obtains (see also \cite{doikou-fioravanti-ravanini})
\be
W_{j{\cal N}}^{(1)} = -i \sqrt{\kappa} \bar \psi_j, ~~~~~W_{j{\cal N}}^{(2)}= \sqrt{\kappa}\bar \psi_j', ~~~~~ \ldots
\ee
and
\be
Z_{{\cal N}}^{(n)}(B, A,t; \lambda)= i L \delta_{n, -1} +\sqrt{\kappa}\int_{A}^B dx\ \psi_j(x,t)\ W^{(n)}_{j{\cal N}}(x,t;\lambda)
\ee

Given the expression for the local charges $\ln(\mathfrak{T})$ ({\ref{gen1}) we expand in powers if $1 \over \lambda$ and the first non trivial conserved quantity is the modified momentum
\be
{\cal I}^{(2)} = \kappa \sum_{j=1}^{{\cal N}-1} \int_{-L}^0 dx\ \Big (\psi_j(x)\ \bar \psi_j'(x) - \psi_j'(x)\ \bar \psi_j(x)  \Big )+ \kappa\sum_{j=1}^{{\cal N}-1}\psi_j(0)\ \bar \psi_j(0)+ \kappa\sum_{j=1}^{{\cal N}-1}\bar \psi_j(0)\ \bar \psi_j(0).
\ee
This result confirms earlier findings on the vector NLS model, which were obtained from the study of the discrete version of the NLS model \cite{doikou-fioravanti-ravanini}. Application of this setting in other physical systems associate to higher rank algebra such as affine Toda field theories or higher rank $\sigma$ models would certainly provide significant physical results, but these issues will be addressed in forthcoming investigations.

\section{Discussion}

We have formulated a classical variation of the quantum vertex algebra, and we have shown that the auxiliary linear function is basically a classical vertex operator. Based on these notions we considered the problem of identifying the local integrals of motion and time components of the corresponding Lax pairs. Particular emphasis was given in systems in presence of defects and non-trivial integrable boundary conditions.  It turns out that in these cases and in particular in the boundary case this formulation is technically considerably simpler and provides results directly at the continuum level even in situations that this has not been possible in a straightforward manner such as e.g. in the study of the vector NLS with twisted Yangian symmetry.

One of course has to note that the choice of the vectors $V,\ V^*$ implemented at the ends of the system modifies or even perhaps restricts the behavior of the system under study. It is worth pointing out however that complications arising due to the asymptotic behavior of the monodromy matrix in systems associated to high rank algebras on the half line can be efficiently dealt with in this framework, basically due to the restrictions imposed from the vectors $V,\ V^*$. In fact, it would be illuminating and would provide further physical insight to implement this frame in prototype classical systems such as affine Toda field theories with integrable boundary conditions, however this issue will be addressed in a separate publication.

At the quantum level and in the context of quantum inverse scattering method in particular one builds the one dimensional lattice system using the ``bare'' $R$-matrix. Then the solution of the Bethe ansatz equations (BAE) in the thermodynamic limit serves as a renormalization process, so one may for instance derive the physical scattering information solving the BAE in the thermodynamic limit. Formulation of a bare version of vertex operators and the corresponding vertex algebra using the bare $R$-matrix could be the first step towards reconciling the Bethe anaszt formalism with the vertex operator construction by the Japanese group \cite{japanese}. The light cone lattice approach by Destri and de Vega \cite{destri-devega} could serve as a possible bridge between these two methodologies. This is a very interesting direction to pursue and will be left for future investigations.
\\
\\
{\bf Acknowledgements}\\
I am indebted to Robert Weston for illuminating discussions on vertex operators as well as useful comments on the manuscript.

\end{document}